\documentclass[10pt,twocolumn,final]{article}
\pdfoutput=1

\usepackage[top=2cm, left=1.5cm, bottom=2cm, right=1.5cm]{geometry}

\usepackage{latexsym}
\usepackage{graphicx}
\usepackage{balance}

\usepackage{array}
\usepackage{multirow}

\usepackage[square,numbers,sort&compress]{natbib}

\usepackage{fancyhdr}
\makeatother
\usepackage{url}
\usepackage{xcolor}
\usepackage{pdfpages}
\usepackage[title]{appendix}

\usepackage{amssymb}
\usepackage{amsmath}
\usepackage{bm}

\newcommand{\fscore}[1]{F\textsubscript{#1}-score}
\newcommand{\fscores}[1]{F\textsubscript{#1}-scores}

\usepackage{caption}
\newcommand{\customCaption}[2]{
\vspace{-15pt}
\begin{flushleft}
\footnotesize
\textbf{Figure~#1}: #2
\end{flushleft}
}

\newcommand{\customCaptionTable}[2]{
\vspace{-15pt}
\begin{flushleft}
\footnotesize
\textbf{Table~#1}: #2
\end{flushleft}
}

\begin{document}
\twocolumn[
  \begin{@twocolumnfalse}

\title{
Fine-Grained Hard Negative Mining: Generalizing Mitosis Detection with a Fifth of the MIDOG 2022 Dataset
}
\date{}

\vspace*{-50pt}
\begin{minipage}{\textwidth}
\centering
\author{
Maxime W. Lafarge and Viktor H. Koelzer
}
\end{minipage}

\maketitle

\vspace*{-20pt}
\begin{center}
\begin{minipage}{0.9\textwidth}
\begin{flushleft}
\footnotesize
\textit{Department of Pathology and Molecular Pathology, University Hospital and University of Zurich, Zurich, Switzerland} 
\end{flushleft}
\end{minipage}
\end{center}

\vspace*{20pt}
\begin{abstract}
Making histopathology image classifiers robust to a wide range of real-world variability is a challenging task.
Here, we describe a candidate deep learning solution for the Mitosis Domain Generalization Challenge 2022 (MIDOG) to address the problem of generalization for mitosis detection in images of hematoxylin-eosin-stained histology slides under high variability (scanner, tissue type and species variability).
Our approach consists in training a rotation-invariant deep learning model using aggressive data augmentation with a training set enriched with hard negative examples and automatically selected negative examples from the unlabeled part of the challenge dataset.

To optimize the performance of our models, we investigated a hard negative mining regime search procedure that lead us to train our best model using a subset of image patches representing $19.6$\% of our training partition of the challenge dataset.
Our candidate model ensemble achieved a \fscore{1} of $.697$ on the final test set after automated evaluation on the challenge platform, achieving the third best overall score in the MIDOG 2022 Challenge.
\end{abstract}
\vspace*{40pt}

  \end{@twocolumnfalse}
]

\section*{Introduction}
To support the research community with the development of new mitosis detection algorithms that are robust to scanner variability, the \textit{MIDOG 2021 Challenge} \cite{midog2021} lead to an overview of efficient approaches towards solving this task.
To further encourage the development of models that can generalize beyond inter-scanner variability, the \textit{MIDOG 2022 Challenge} was initiated \cite{midog2022description},
offering a unique opportunity to compare the generalization ability of mitosis detectors in a blind manner, as the challenge organizers independently evaluate candidate solutions on held-out sets of images from undisclosed and unseen domains.

This opportunity motivated us to revisit the pioneer methodology proposed by Cire{\c{s}}an et al. \citep{cirecsan2013mitosis}, and to assess the relative performance of standard methods in the context of modern deep convolutional neural network architectures and large high-variability mitosis datasets as the one provided for this challenge.

Our approach consists of a patch-based training procedure that we used to train deep learning models that can then be applied to detect mitoses on unseen images, building upon the strategy we employed for the \textit{MIDOG 2021 challenge} \citep{lafarge2021midog}.
We incrementally made changes to this training procedure over the development phase of the challenge and monitored performances on our validation partition of the challenge dataset in order to select and submit our best performing model.
In this paper, we describe the different components that constitute our submitted solution, including a fine-grained assessment of a hard negative mining procedure that we consider to be the main contributing part of our solution.

\begin{figure}[h!]
\begin{center}
\includegraphics[width=\columnwidth, trim=5pt 5pt 5pt 5pt, clip]{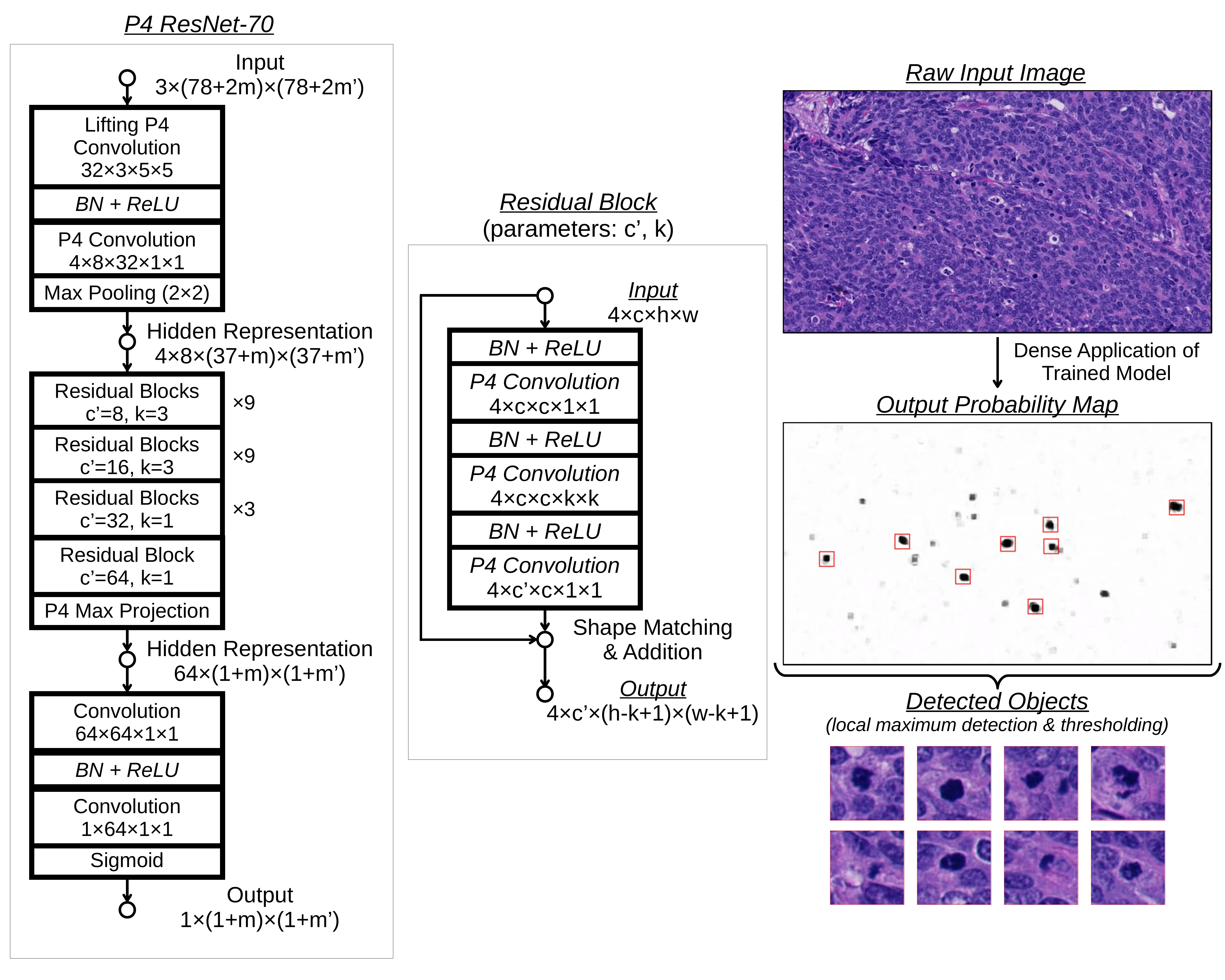}
\caption{} 
\label{fig:modelArchitecture}
\customCaption{\ref{fig:modelArchitecture}}{
Architecture of the 70-layer ResNet used in this work and example of its application to a raw input H\&E histopathology image.
The shape of output tensors is written with the following format: $\textrm{(\textit{Orientations}}{\times}\textrm{)}\textrm{\textit{Channels}}{\times}\textrm{\textit{Height}}{\times}\textrm{\textit{Width}}$.
The shape of trainable operator tensors is written with the following format: $\textrm{(\textit{Orientations}}{\times}\textrm{)}\textrm{\textit{Out.Channels}}{\times}\textrm{\textit{In.Channels}}{\times}\textrm{\textit{Ker.Height}}{\times}\textrm{\textit{Ker.Width}}$.
}
\end{center}
\end{figure}

\vspace{10pt}
\begin{table*}[ht!]
\begin{center}
\caption{} 
\label{tab:dataAugmentation}
\customCaptionTable{\ref{tab:dataAugmentation}}{
Data augmentation protocol: for each input image patch, the following list of transformations is scanned and applied with a given probability.
Transformation parameters are randomly sampled in a given interval.
The two variants of the protocol used to train our submitted model ensemble are detailed here.
}

\renewcommand{\arraystretch}{1.2}
\footnotesize
\begin{tabular}{p{0.5\columnwidth} p{0.25\columnwidth} p{0.2\columnwidth} p{0.25\columnwidth} p{0.2\columnwidth}}
\multirow{2}{*}{\textit{Transformation}}
& \multicolumn{2}{c}{\textit{Policy A}}
& \multicolumn{2}{c}{\textit{Policy B}}
\\

{\ }
& \textit{Coefficients}
& \textit{Probability} 
& \textit{Coefficients}
& \textit{Probability} 
\\\hline\hline

Transposition & -- & $50\%$ & -- & $50\%$ \\\hline
Elastic Deformation & -- & $100\%$ & -- & $50\%$ \\\hline

Spatial Shift ($\Delta_{x,y}$) & $\left[-12\text{px}, 12\text{px}\right]$ & $100\%$ & $\left[-12\text{px}, 12\text{px}\right]$ & $100\%$ \\\hline
Spatial Zoom ($\alpha$) & $\left[-10\%, 20\%\right]$ & $50\%$ & $\left[-10\%, 20\%\right]$ & $50\%$ \\\hline

(HLS) Hue Rotation ($h$) & $\left[0^{\circ}, 360^{\circ}\right]$ & $80\%$ & $\left[-60^{\circ}, 60^{\circ}\right]$ & $50\%$ \\\hline
Color Shift ($c_{r,g,b}$) & $\left[-51, 51\right]$ & $80\%$ & $\left[-51, 51\right]$ & $50\%$ \\\hline
Contrast Correction ($\mu_{r,g,b}$) & $\left[0.8, 1.2\right]$ & $80\%$ & $\left[0.8, 1.2\right]$ & $50\%$ \\\hline
Gamma Correction ($\gamma_{r,g,b}$) & -- & $0\%$ & $\left[0.8, 1.2\right]$ & $50\%$ \\\hline

\end{tabular}
\end{center}
\end{table*}

\section*{Model Architecture}
We implemented a customized $70$-layer ResNet architecture \citep{he2016preActResNet} to model the confidence probability for input images to be centered on a mitotic figure within a receptive field of $78{\times}78$ pixels.
We replaced standard convolutional layers by P4-group convolutional layers \citep{cohen2016groupCNN} to guarantee invariance of our models to $90$-degree rotations without requiring train-time or test-time rotation augmentations, with a low computational overhead.
This change was further motivated by the improvment of performance across multiple histopathology image classification tasks for models using this type of operation reported in the literature \citep{veeling2018rotation,lafarge2020roto,graham2020dense}.
This architecture was adjusted to enable dense application of the models to arbitrarily large input images.
A detailed flowchart of this architecture and an example of the application of a trained model to an input image are shown in Figure~\ref{fig:modelArchitecture}.

\section*{Dataset Partitioning}
To train models and evaluate their performance, we exclusively used the data provided for the track $1$ of the \textit{MIDOG 2022 Challenge} \cite{midog2022description}.
We split the provided annotated images according to a $80$-$20$ training-validation scheme such that labels and domains were stratified (training set: $7588$ mitoses from $283$ images; validation set: $1913$ mitoses from $71$ images).

Given the provided ground-truth locations of mitotic figures in these images,
we derived a set of image patches of positive examples centered on mitotic figures
and a set of image patches of all negative examples whose center is sufficiently distant from the center of annotated mitotic figures.

\section*{Training Procedure}
All models were trained by minimizing the cross-entropy loss via stochastic gradient descent with momentum (initial learning rate $0.03$ and momentum $0.9$) using input mini-batches of size $128$.
We used a cyclic learning rate scheduling \cite{loshchilov2017sgdr} with a cycle period of $10$k iterations and applied weight decay regularization (coefficient $10^{-4}$).
Mini-batches were generated with randomly sampled image patches of size $78{\times}78$, balanced between positive and negative examples.
All image patches were randomly transformed according to an augmentation protocol (including channel-wise intensity distortions) whose operations and parameter ranges are detailed in Table.~\ref{tab:dataAugmentation}.
For evaluation purposes, we saved the weights of the model that achieved the lowest validation loss within $150$k training iterations.

\begin{figure}[h!]
\begin{center}
\includegraphics[width=\columnwidth]{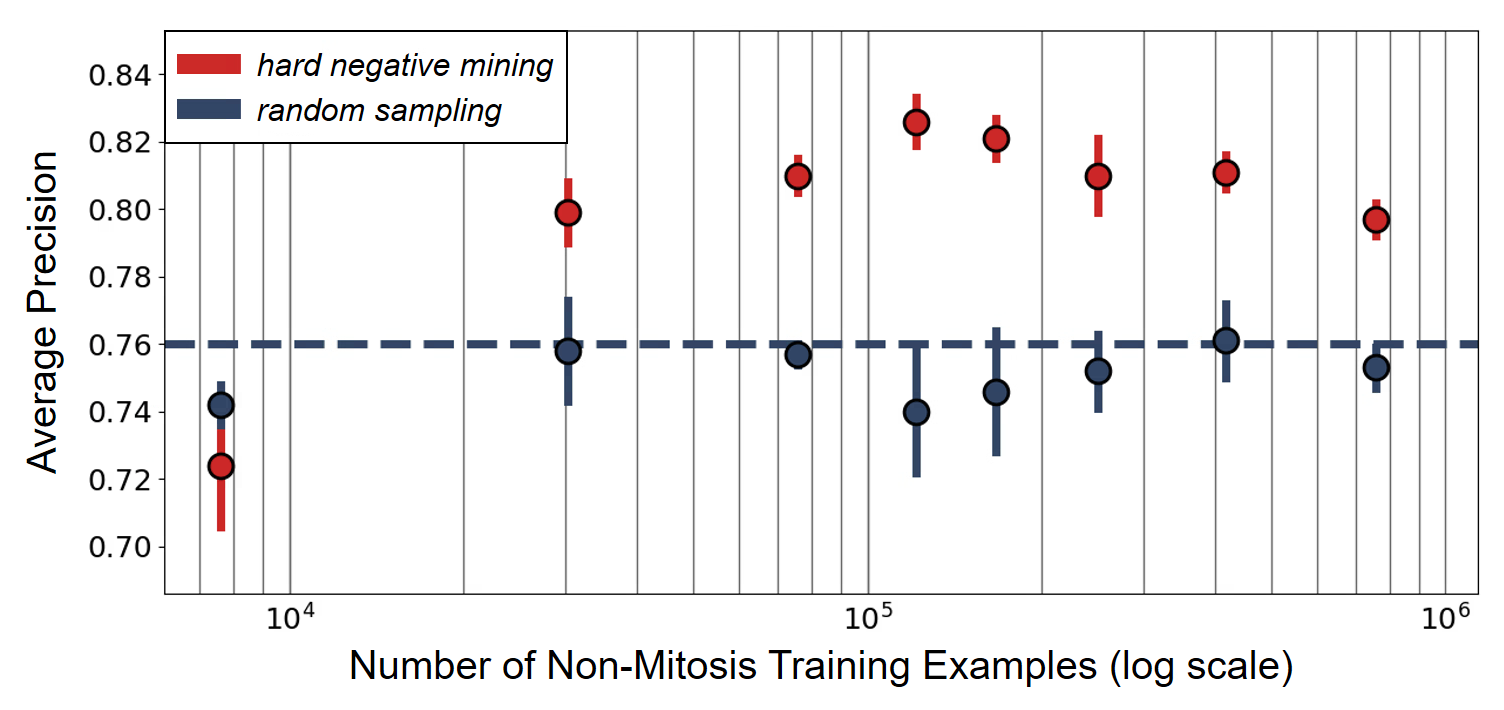}
\caption{} 
\label{fig:hnmRegime}
\customCaption{\ref{fig:hnmRegime}}{
Average precision (AP) on the validation set of models trained with subsets of non-mitosis examples of different sizes.
These subsets were either generated via random sampling or by hard negative mining.
During training, the positive class is oversampled to ensure balance of classes in training batches.
Circles and bars represent the mean and standard deviation of AP for three repeated experiments with different random seeds.
The dashed line indicates the mean AP obtained when training using all the possible negative examples of our training partition.
}
\end{center}
\end{figure}

\section*{Fine-Grained Hard Negative Mining}
Hard negative mining (HNM) has become a standard procedure for the development of mitosis detectors since the solution proposed by \cite{cirecsan2013mitosis}, and aims at improving the model performance by using a well-chosen subset of "hard" negative examples for training instead of using randomly sampled negative examples.
This procedure typically requires training a first model with all the available positive and negative examples,
then ranking all negative examples based on the confidence score output by this trained model,
and finally keeping the negative examples with a score above a fixed cutoff threshold as a set of "hard negatives" to be used to repeat training and improve performances.

For this challenge, we considered this threshold as a hyper-parameter and searched for an optimal number of hard negative examples to select that would maximize performance on the validation set.
A summary of the binary search we conducted for this parameter is shown in Figure~\ref{fig:hnmRegime}.

This fine-grained assessment enabled the selection of an optimal subset of $121~738$ hard negative examples of our training partition, which improved validation performances in comparison to using randomly sampled training examples.
Our submitted model ensemble was trained using this optimal subset of image patches along with all positive examples whose joint total pixel count represents $19.6$\% of the overall pixel count of our training partition.

\vspace{10pt}
\begin{table*}[ht!]
\begin{center}
\caption{} 
\label{tab:scores}
\customCaptionTable{\ref{tab:scores}}{
Comparison of \fscores{1} of our trained models (two different augmentation policies and their ensemble) and baselines of the \textit{MIDOG 2022 Challenge} on the four tumor types (\textit{T1,2,3,4}) of the hidden preliminary test set and on the final test set of the challenge.
}

\newlength{\scoreLen}
\setlength{\scoreLen}{0.2\textwidth}

\newcommand{\cc}[1]{\multicolumn{1}{c}{\hspace{5pt} #1 \hspace{5pt}}}
\renewcommand{\arraystretch}{1.2}
\footnotesize
\begin{tabular}{m{0.15\textwidth} m{\scoreLen} m{1pt} m{\scoreLen} m{\scoreLen} m{\scoreLen} m{\scoreLen} m{\scoreLen} m{1pt} m{\scoreLen}}
   \multirow{2}{*}{\textit{Model}}
&  \cc{\textit{Internal}}
&  
&  \multicolumn{5}{c}{\textit{Preliminary Test}}
&  
&  \multicolumn{1}{c}{\textit{Final Test}}
\\

   \cc{\textit{}}
&  \cc{\textit{Validation}}
&
&  \cc{\textit{Overall}}
&  \cc{\textit{T1}}
&  \cc{\textit{T2}}
&  \cc{\textit{T3}}
&  \cc{\textit{T4}}
&
&  \cc{\textit{Overall}} \\\hline\hline

ours \newline (\textit{Policy A})
& \cc{$.784$} 
&  
& \cc{$.646$}
& \cc{$\mathbf{.783}$}
& \cc{$.726$}
& \cc{$.548$}
& \cc{$.708$} 
&  
& \cc{--}
\\\hline

ours \newline (\textit{Policy B})
& \cc{$.787$} 
& 
& \cc{$.571$}
& \cc{$.757$}
& \cc{$\mathbf{.775}$}
& \cc{$.428$}
& \cc{$.571$}
& 
& \cc{--}
\\\hline

ours \newline (Ensemble)
& \cc{$\mathbf{.791}$} 
& 
& \cc{$.690$}
& \cc{$.758$}
& \cc{$.735$}
& \cc{$.653$}
& \cc{$.610$}
& 
& \cc{$.696$}
\\\hline\hline

MIDOG 2022 \newline (Baseline 2) 
& \cc{--}
& 
& \cc{$\mathbf{.715}$}
& \cc{$.744$}
& \cc{$.732$}
& \cc{$\mathbf{.692}$}
& \cc{$.711$}
& 
& \cc{$\mathbf{.714}$}
\\\hline

MIDOG 2022 \newline (Baseline 1)
& \cc{--}
& 
& \cc{$.629$}
& \cc{$.753$}
& \cc{$.608$}
& \cc{$.585$}
& \cc{$\mathbf{.743}$}
& 
& \cc{--}
\\\hline

\end{tabular}
\end{center}
\end{table*}

\section*{Automated Enrichment\newline with Negative Examples}
\begin{figure}[h!]
\begin{center}
\includegraphics[width=\columnwidth]{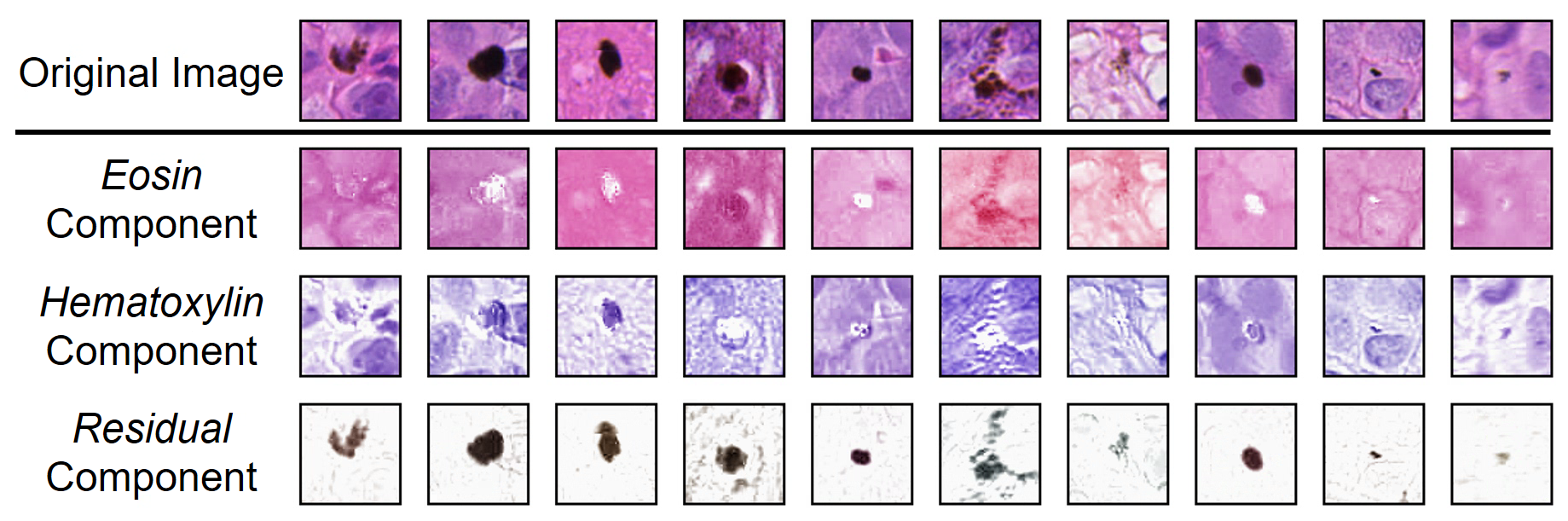}
\caption{} 
\label{fig:residualScreening}
\customCaption{\ref{fig:residualScreening}}{
Example of automatically selected image patches from the unlabeled part of the \textit{MIDOG 2022} dataset based on the optical density of their residual component after application of a stain unmixing procedure.
These selected image patches were used to enrich the training set.
}
\end{center}
\end{figure}

To further enrich the dataset with additional negative examples we implemented a stain unmixing algorithm derived from \cite{macenko2009method} to first separate the hematoxylin, eosin and residual components in the unlabeled images of the dataset,
and then automatically extract image patches with high optical density in the estimated residual component to enrich the training set with extra negative examples.
This procedure resulted in the selection of $72$ negative examples after application of an optimal hard negative mining selection rule as described in the previous section. A batch of such selected examples is shown in Figure~\ref{fig:residualScreening}.
This approach was motivated by the observation that true positive mitotic events reside in defined stain vectors and are separable from background events such as pigmentation and ink that form common impostors.

\vspace{10pt}

\section*{Inference Pipeline}
Once our models were trained, we produced prediction maps by applying them densely on test images, and then derived candidate detection locations as the set of local maxima.
We then considered all candidate locations with a prediction score above a threshold value as final detection locations (this threshold was chosen as the one maximizing the \fscore{1} on the validation set).

For our final submission, we created a model ensemble by taking the agreement between the detection sets of two models trained under two variants of our augmentation protocol as detailed in Table~\ref{tab:dataAugmentation}.
A comparison of performance of our models against baselines is summarized in Table~\ref{tab:scores}.

\section*{Discussion}
We present a candidate algorithm that achieves moderate generalization performance with an overall \fscore{1} of $.696$ on the final test set of the challenge (set of images from ten unseen domains), while relying mostly on conventional methods (patch-based training procedure, ResNet architecture, cross-entropy loss minimization, hard negative mining, standard augmentation protocol).

Performance on the different domains of the test sets were heterogeneous indicating that the investigated pipeline only enables generalization to a subset of the held-out test domains of the challenge.

Hard Negative Mining played an important role in achieving the final performance of our submitted solution: the search for an optimal amount of hard negative training examples helped improving performances on the validation set with an increase of Average Precision from $.760$ to $.826$.
Our comparative analysis suggests that this level of performance could not have been achieved from training using only arbitrary negative examples.

Furthermore, we investigated the use of a conventional stain unmixing method to automatically extract potential negative examples in unlabeled images.
As this approach did not decrease the performances on the validation set, we assumed this helped our models better generalizing.
We thus used the resulting enriched training set to train our final models, yet, because of the very small number of training examples selected this way, we cannot further conclude whether the effect of this strategy was beneficial or not regarding the performance on the final test set.
Because this enrichment procedure relies on the strong assumption that extracted examples belong to the non-mitosis class, it should be kept in mind that positive examples could still potentially be extracted.
Here, we recommend that the examples extracted by such a method should ideally be reviewed by expert annotators to ensure the correctness of the class of these additional training examples.

Although the reported variations of our augmentation protocol produced similar results on the validation set, we observed that they produced heterogeneous performances on the preliminary test set (Table~\ref{tab:scores}).
This suggests that generalization to the domains of the preliminary test set is sensitive to variations of the augmentation protocol used for training.
Indeed, the two reported augmentation policies (Table~\ref{tab:dataAugmentation}) helped generalizing to different parts of the unseen variability of the preliminary test set (\textit{Policy A} enabled better generalization for Tumor $1$ whereas \textit{Policy B} enabled better generalization for Tumor $2$).
These results corroborate the known association between chosen augmentation policies and generalization performance: this motivates us to further study how to best design and select augmentation protocols to improve domain generalization.


\subsection*{References}
\balance
\vspace*{20pt}
\bibliographystyle{unsrtnat}
\renewcommand\refname{}
\vspace*{-40pt}
\footnotesize
\bibliography{main}

\newpage
\appendix


\end{document}